\documentclass[a4paper]{article}

\usepackage{18lomcon}        
\usepackage{cite}             
\usepackage{hyperref,bm,amsmath,graphicx,esint,amssymb,subfigure}

\begin{document}


\title{MAGNETIC FIELD GENERATION IN DENSE GAS OF MASSIVE ELECTRONS WITH ANOMALOUS MAGNETIC MOMENTS ELECTROWEAKLY INTERACTING WITH BACKGROUND MATTER}

\author{Maxim Dvornikov \email{maxdvo@izmiran.ru}}

\affiliation{Pushkov Institute of Terrestrial Magnetism, Ionosphere and Radiowave Propagation (IZMIRAN), 108840 Troitsk, Moscow, Russia}

\affiliation{Physics Faculty, National Research Tomsk State University, 36 Lenin Avenue, 634050 Tomsk, Russia}


\date{\today}
\maketitle


\begin{abstract}
We describe the generation of the electric current flowing along the external magnetic field in the system of massive charged fermions, possessing anomalous magnetic moments and electroweakly interacting with background matter. This current is shown to result in the instability of the magnetic field leading to its growth. Some astrophysical applications are discussed.
\end{abstract}

Strong magnetic fields with $B > 10^{15}\,\text{G}$ can be found in some compact stars called magnetars~\cite{TurZanWat15}. Recently the elementary particle physics mechanisms, involving the chiral magnetic effect (CME)~\cite{Vil80}, were applied to model magnetic fields in magnetars~\cite{Kha15}. CME is based on the existence of the current $\mathbf{J} = \alpha_{\mathrm{em}} \left( \mu_{\mathrm{R}}-\mu_{\mathrm{L}} \right) \mathbf{B}/\pi$ in the external magnetic field $\mathbf{B}$. Here $\alpha_{\mathrm{em}} = e^2/4\pi$ is the fine structure constant, $e>0$ is the absolute value of the elementary charge, and $\mu_{\mathrm{R,L}}$ are the chemical potentials of right and left chiral fermions,. However CME implies the unbroken chiral symmetry~\cite{Dvo16a}, i.e. charged particles should be considered massless. The electroweak chiral phase transition is unlikely in astrophysical media. There is a possibility to restore chiral symmetry in quark matter owing to QCD effects. Thus one can expect the magnetic field amplification in the core of some compact stars where quark matter can be present~\cite{Dvo16b}.

In this paper, we summarize our recent findings in~\cite{Dvo17} on the magnetic field generation in the system of massive electrons with nonzero anomalous magnetic moments. We shall suppose that, besides the external magnetic field, these electrons electroweakly interact with background matter. It turns out that there is a nonzero electric current of these electrons along the magnetic field. The computation of the current is based on the exact solution of the Dirac equation in the considered external fields which was recently found in~\cite{BalStuTok12}. If the current $\mathbf{J} \sim \mathbf{B}$ is taken into account in the Maxwell equations, it will result in the magnetic field instability causing the enhancement of a seed field. Finally, we briefly discuss some astrophysical applications.

Let us consider a massive fermion, i.e. an electron, having the anomalous magnetic moment $\mu$. We suppose that this electron electroweakly interacts with background matter under the influence of the external magnetic field. We found in~\cite{Dvo17} that, in this situation, there is a nonzero electric current along the magnetic field,
\begin{equation}\label{eq:JPiB}
  \mathbf{J} = \Pi\mathbf{B},
  \quad
  \Pi=-8\mu mV_{5}B\frac{\alpha_{\mathrm{em}}}{\pi\tilde{\chi}^{3}}
  \sum_{\mathrm{n}=1}^{N}
  \sqrt{\tilde{\chi}^{2}-m_{\mathrm{eff}}^{2}},
\end{equation}
where $m$ is the electron mass, $V_5 = (V_{\mathrm{L}}-V_{\mathrm{R}})/2$, $V_{\mathrm{R,L}}$ are the effective potentials of the electroweak interaction for right and left chiral projections, which are given in~\cite{DvoSem15a}, $m_{\mathrm{eff}}=\sqrt{m^{2}+2eB\mathrm{n}}$, $\mathrm{n} = 0,1,\dots$ is the discrete quantum number which energy levels depend on, $\tilde{\chi}=\chi-\bar{V}$, $\chi$ is the chemical potential, $\bar{V} = (V_{\mathrm{L}}+V_{\mathrm{R}})/2$, and $N$ is maximal integer, for which $\tilde{\chi}^{2}-m^{2}-2eBN\geq0$. In Eq.~\eqref{eq:JPiB}, we consider highly degenerate electron gas, in which $\chi \gg T$, where $T$ is the temperature.

If we take into account the current in Eq.~\eqref{eq:JPiB} in the Maxwell equations and consider the MHD approximation, the amplitude of the magnetic field obeys the equation,
\begin{equation}\label{eq:dotB}
  \dot{B}=-\frac{k}{\sigma_{\mathrm{cond}}}
  \left(
    k+\Pi
  \right)B,
\end{equation}
where $\sigma_{\mathrm{cond}}$ is the electric conductivity and the parameter $k$ determines the length scale $L=1/k$ of the magnetic field. Since $\Pi$ in Eq.~\eqref{eq:JPiB} is negative, the magnetic field,
described by Eq.~\eqref{eq:dotB}, can be unstable.

\begin{figure}
  \centering
  \subfigure[]
  {\label{1a}
  \includegraphics[scale=.09]{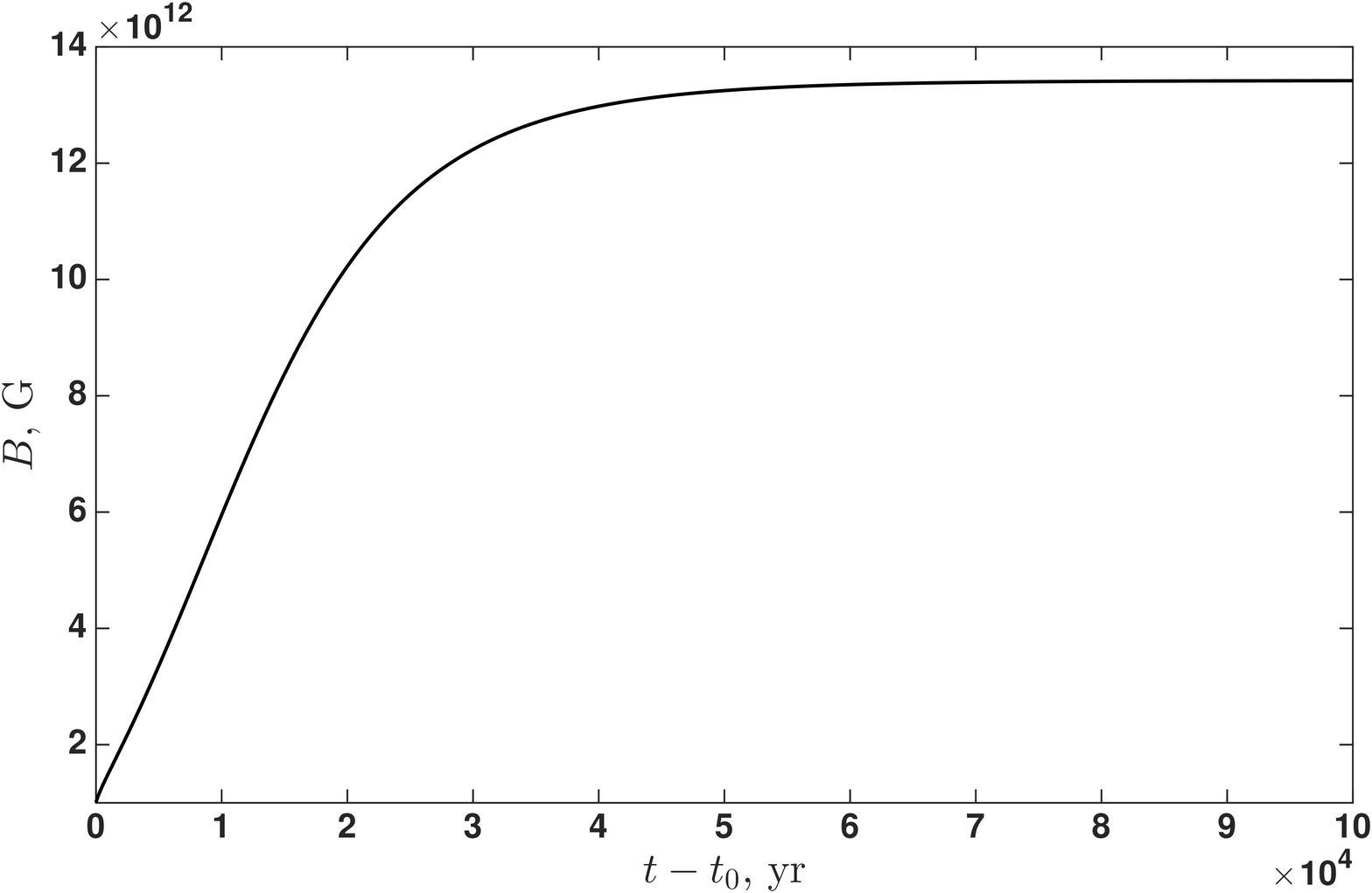}}
  \hskip-.75cm
  \subfigure[]
  {\label{1b}
  \includegraphics[scale=.09]{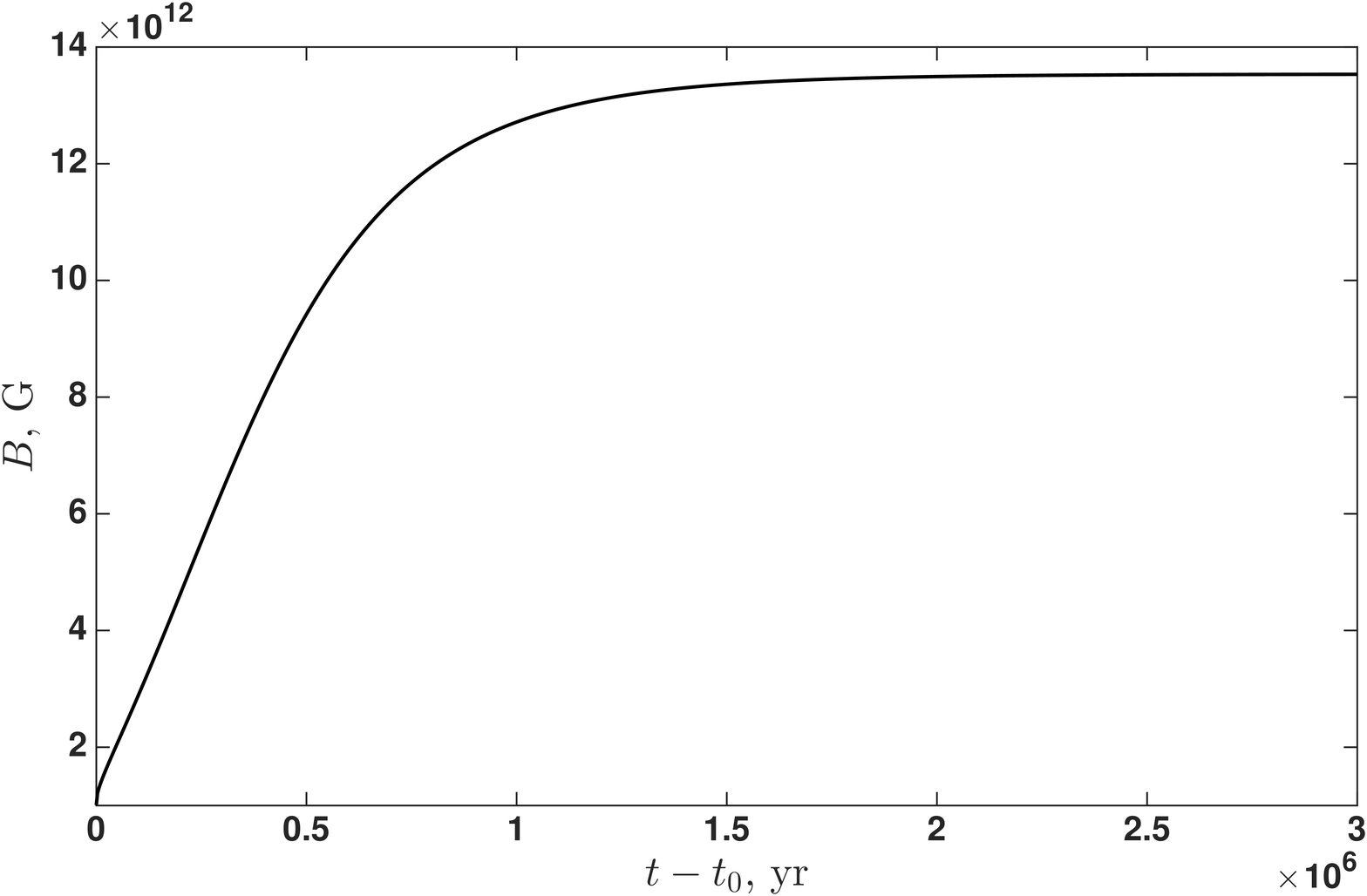}}
  \protect
  \caption{Magnetic field evolution obtained in the numerical solution of  
  Eq.~\eqref{eq:dotB} for different length scales.
  (a) $L=10^{2}\,\text{cm}$, and (b) $L=10^{3}\,\text{cm}$.
  \label{fig:Bfield}}
\end{figure}

Basing on Eq.~\eqref{eq:dotB}, we can describe the magnetic field
amplification in a dense degenerate matter which can be found in a neutron star.
Taking into account the typical density of matter in a neutron star, the time dependence of the electric conductivity~\cite{DvoSem15b} and the dependence of $\mu$ on the magnetic field~\cite{Ter69}, in Fig.~\ref{fig:Bfield}, we plot the time dependence of the magnetic field for different length scales starting from the seed magnetic field $B_{0}=10^{12}\,\text{G}$, which is typical for a young pulsar.

One can see that the magnetic reaches the saturated magnetic field $B_{\mathrm{sat}} \approx 1.3\times10^{13}\,\text{G}$. The strength of $B_{\mathrm{sat}}$ and its length scale are close to the values of the corresponding parameters in a magnetic field fluctuation~\cite{Dvo16c}, which then can trigger the burst of a magnetar~\cite{TurZanWat15}. We have suggested in~\cite{Dvo17} that the energy source, powering the magnetic field growth presented in Fig.~\ref{fig:Bfield},
can be the kinetic energy of the stellar rotation.

Finally it is interesting to compare the appearance of the new current along the magnetic field in Eq.~\eqref{eq:JPiB} with the CME~\cite{Vil80}. As shown in~\cite{Vil80}, only massless fermions at the zero Landau level in an external magnetic field contribute to the generation of the anomalous current along the magnetic field since there is an asymmetry in the motion of these particles with respect to the external magnetic field~\cite{DvoSem15a,DvoSem15b}. In the situation described in the present work, i.e. when massive electrons with nonzero anomalous magnetic moment propagate in the electroweak matter, the particles at the lowest energy level can move in any direction with respect to the magnetic field without an asymmetry. On the contrary, higher energy levels with $\mathrm{n} > 0$ are not symmetric for electrons moving along and opposite $\mathbf{B}$~\cite{Dvo17}. Moreover, the term in the energy spectrum responsible for such an asymmetry is proportional to $\mu B m V_5$~\cite{BalStuTok12}. It is this factor, which $\mathbf{J}$ in Eq.~\eqref{eq:JPiB} depends on.

\section*{Acknowledgments}

I am thankful to the Tomsk State University Competitiveness Improvement Program and to RFBR (research project No.~15-02-00293) for a partial support.


\end{document}